\documentclass[11pt]{article}
\usepackage{moriond,epsfig}

\def\be{\begin{equation}}
\def\ee{\end{equation}}
\def\bea{\begin{eqnarray}}
\def\eea{\end{eqnarray}}

\begin{document}
\vspace*{4cm}
\title{Forward-backward asymmetry of $B\to K_J l^+l^-$ decays in SM and new physics models}

\author{ Cai-Dian L\"u$^a$ and Wei Wang$^b$}

\address{$^a$ Institute  of  High  Energy  Physics
Chinese Academy of Sciences, P.O. Box 918, Beijing 100049,  China
\\
$^b$ Deutsches Elektronen-Synchrotron DESY, Hamburg 22607, Germany}

\maketitle \abstracts{ We report on  our   studies of $B\to K_Jl^+l^-$ in the standard model and several   new physics variations, with $K_J$ denoting a kaonic resoanance.
   In terms of helicity amplitudes, we derive a compact form for the full angular distributions,
 and use them to calculate the branching ratios, forward-backward asymmetries and polarizations.
 We have  updated the
constraints on effective Wilson coefficients and/or free parameters
in these  new physics scenarios by making use of the $B\to
K^*l^+l^-$ and $b\to s l^+l^-$ experimental data. Their impact on
$B\to K_J^*l^+l^-$ is subsequently explored and in particular the
 zero-crossing point for  forward-backward asymmetry in  new
physics scenarios can sizably deviate from the standard model.   }

\section{Introduction}

Flavor changing neutral currents are forbidden at tree level in
the standard model (SM).
Such rare B-decays into dileptons are precision probes of the  SM and provide constraints on new physics beyond the standard model~\cite{Ali:1999mm}. Important semileptonic modes in terms of experimental accessibility and theory control are those into  $K_J$~\cite{k0,k1,k2,kj,zprime}, in which  $K_J$ can be $K^*$, $K_{1}(1270), K_1(1410)$,
 $K^*(1410), K_0^*(1430), K_2^*(1430), K^*(1680), K_3^*(1780)$ and $K_4^*(2045)$. These decays exhibit a rich phenomenology   through the angular analysis of subsequent decays of $K_J$, through which the
forward-backward asymmetry (FBA) can be extracted. As opposed to  the branching ratios which suffer from large hadronic uncertainties, the  FBA is theoretically clean and sensitive to NP. Therefore it is one of the
major goals of LHCb to precisely explore FBA as a hunt for new physics signals~\cite{Aaij:2011aa}.

\section{Form factor relations}

The $B\to K^*_J$ form factors are nonperturbative in nature and the
application of QCD theory to them  mostly resorts to the Lattice QCD
simulations, which has large  limitation at the current stage.  An  important observation is that,
 in the heavy
quark limit $m_b\to \infty$ and the large energy limit $E\to
\infty$, interactions of the heavy and light systems can be expanded
in small ratios $\Lambda_{QCD}/E$ and $\Lambda_{QCD}/m_B$.
 At the leading
power in $1/m_b$, the large energy symmetry is obtained and  such symmetry greatly simplifies the heavy-to-light
transition~\cite{hep-ph/9812358}.

\begin{table}
\caption{  $B\to K_J^{*}$ form factors derived from the large recoil symmetry. }
 \begin{center}
 \begin{tabular}{ c c c ccc c c}
\hline \hline
 $K_J^{*}$       & $\xi_{||}$  & $\xi_{\perp}$    \\
 \hline
\hline
 $K^*(1410)$    & $ 0.22\pm 0.03$ & $0.28\pm 0.04$ \\ \hline
 $K^*_0(1430)$    & $ 0.22\pm 0.03$ & --  \\ \hline
 $K^*_2(1430)$    & $0.22\pm 0.03$ &$0.28\pm 0.04$  \\ \hline
 $K^*(1680)$    & $ 0.18\pm 0.03$ & $0.24\pm 0.05$ \\
 \hline
 $K^*_3(1780)$    & $0.16\pm 0.03$ & $0.23\pm 0.05$ \\ \hline
 $K^*_4(2045)$    & $0.13\pm 0.03$ & $0.19\pm 0.05$ \\ \hline
 \end{tabular}\label{Tab:formfactors}
 \end{center}
 \end{table}

 As a concrete
application,  the current $\bar s\Gamma b$ in QCD can be matched
onto the current $\bar s_n \Gamma b_v$ constructed in terms of the
fields in the low-energy effective theory. Here $v$ denotes the velocity of the
heavy meson and $n$ is a light-like vector along the $K^*_J$ moving
direction. This procedure constrains the independent Lorentz
structures and reduces the seven independent hadronic form factors
for each $B\to K^*_J$ ($J\ge1$)  type to two universal functions
$\xi_\perp$ and $\xi_{||}$. Explicitly, we have
\begin{eqnarray}
 A_0^{K^*_J}(q^2) \left(\frac{|\vec p_{K^*_J}|}{m_{K^*_J}}\right)^{J-1}&\equiv& A_0^{K^*_J, \rm eff}\simeq (1-\frac{m_{K^*_J}^2}{m_B E}) \xi_{||}^{K^*_J}(q^2) +\frac{m_{K^*_J}}{m_B} \xi_{\perp}^{K^*_J}(q^2),\nonumber\\
 A_1^{K^*_J}(q^2) \left(\frac{|\vec p_{K^*_J}|}{m_{K^*_J}}\right)^{J-1}&\equiv& A_1^{K^*_J, \rm eff}\simeq
 \frac{2E}{m_B+m_{K^*_J}} \xi_{\perp}^{K^*_J}(q^2),\nonumber\\
 A_2^{K^*_J}(q^2) \left(\frac{|\vec p_{K^*_J}|}{m_{K^*_J}}\right)^{J-1}&\equiv& A_2^{K^*_J, \rm eff}\simeq
 (1+\frac{m_{K^*_J}}{m_B})[\xi_{\perp}^{K^*_J}(q^2) -\frac{m_{K^*_J}}{E}\xi_{||}^{K^*_J}(q^2)],\nonumber\\
V^{K^*_J}(q^2) \left(\frac{|\vec
p_{K^*_J}|}{m_{K^*_J}}\right)^{J-1}&\equiv& V^{K^*_J, \rm eff}\simeq
 (1+\frac{m_{K^*_J}}{m_B}) \xi_{\perp}^{K^*_J}(q^2),\nonumber\\
T_1^{K^*_J}(q^2) \left(\frac{|\vec p_{K^*_J}|}{m_{K^*_J}}\right)^{J-1}&\equiv& T_1^{K^*_J, \rm eff}\simeq\xi_{\perp}^{K^*_J}(q^2),\nonumber\\
T_2^{K^*_J}(q^2) \left(\frac{|\vec
p_{K^*_J}|}{m_{K^*_J}}\right)^{J-1}&\equiv& T_2^{K^*_J, \rm
eff}\simeq
 (1-\frac{q^2}{m_B^2-m_{K^*_J}^2}) \xi_{\perp}^{K^*_J}(q^2),\nonumber\\
 T_3^{K^*_J}(q^2) \left(\frac{|\vec p_{K^*_J}|}{m_{K^*_J}}\right)^{J-1}&\equiv& T_3^{K^*_J, \rm eff}\simeq
\xi_{\perp}^{K^*_J}(q^2) - (1-\frac{m_{K^*_J}^2}{m_B^2})
\frac{m_{K^*_J}}{E} \xi_{||}^{K^*_J}(q^2).\label{eq:LEETrelation}
\end{eqnarray}
In the case of $B$ to scalar meson
transition, the large energy limit gives
\begin{eqnarray}
 \frac{m_{B}}{m_{B}+m_{K^*_0}} F_T(q^2) = F_1(q^2) = \frac{m_B}{2E} F_0(q^2)=\xi^{K^*_0}(q^2).
\end{eqnarray}
The results for $\xi_{||}^{K^*_J}$ and $\xi_{\perp}^{K^*_J}$ obtained
from the Bauer-Stech-Wirbel (BSW) model~\cite{Wirbel:1985ji} in
Ref.~\cite{Hatanaka:2009sj} are used in our work and we collect
these results in Tab.~\ref{Tab:formfactors}. For the $B\to K^*_0$
transition,  it is plausible to employ $\xi^{B\to K_0^*}=
\xi_{||}^{B\to K_2^*}$ since both $K^*_0$ and $K^*_2$ are p-wave
states.

\begin{table}
\caption{$B\to K^*_2$ form factors at  $q^2=0$ in the ISGW2 model, the
covariant light-front quark model
and the light-cone QCD sum rules and perturbative QCD approach.}
\begin{center}
\begin{tabular}{cccccccc}
\hline \hline
 \hline &
 ISGW2     & CLFQM    & LCSR  & LEET+BSW  & PQCD  \\\hline
  $V^{BK_2^*}$ &  0.38  & 0.29 & $0.16\pm 0.02$ & $0.21 \pm 0.03$ & $0.21^{+0.06}_{-0.05}$ \\
      $A_0^{BK_2^*}$                  &$0.27$ & 0.23 & $0.25\pm 0.04$ & $0.15\pm 0.02$ &$0.18^{+0.05}_{-0.04}$                 \\
       $A_1^{BK_2^*}$                 & 0.24 & 0.22 & $0.14\pm 0.02$ & $0.14\pm 0.02$ & $0.13^{+0.04}_{-0.03}$            \\
      $A_2^{BK_2^*}$              &$0.22$ & 0.21 & $0.05\pm 0.02$ &$0.14\pm 0.02$ & $0.08^{+0.03}_{-0.02}$               \\
        $T_1^{BK_2^*} $     &
             &$0.28$               & $0.14\pm 0.02$  & $0.16\pm 0.02$     &$   0.17_{  -0.04         }^{+   0.05        }
 $      \\
        $T_3^{BK_2^*}$                & &$-0.25$   & $0.01^{+0.02}_{-0.01}$ & $0.10\pm 0.02$  &$   0.14
            _{  -0.03         }
            ^{+   0.05        }
 $            \\
\hline \end{tabular}\label{Tab:BtoK2formfactorcomparision}
\end{center}
\end{table}

In addition, we have employed the perturbative QCD approach to
directly compute these form
factors~\cite{Li:2008tk,Li:2009tx,Wang:2010ni} and  find many
agreements with the large recoil symmetries, for instance the PQCD
results for $B\to K_2^*$ transition are shown in
 Table~\ref{Tab:BtoK2formfactorcomparision} (See Ref.~\cite{kj} for a more detailed comparison).

\section{New physics contributions}

We choose several kinds of new physics models, such as family
non-universal Z' model, Supersymmetric model and vector-like quark
model. All of them can induce extra contributions to the branching
ratios, polarizations and forward-backward asymmetry parameters,
through the effective operators $O_9$ and/or $O_{10}$. Via modifying
the Wilson coefficients $C_9$ and $C_{10}$, these contributions
affect the observables in $B\to K^*l^+l^-$ as well and the
comparison of theory with data derive the constraints on $C_9$ and
$C_{10}$.

We adopt a least $\chi^2$-fit method and make use of the experimental data on $B\to K^*l^+l^-$.
Embedded in the vector-like quark model, the free two parameters, real part and imaginary
part of the FCNC coupling  $\lambda_{sb}$, are found as
\begin{eqnarray}
 {\rm Re}\lambda_{sb}=(0.07\pm0.04)\times 10^{-3},\;\;\;
 {\rm Im}\lambda_{sb}=(0.09\pm0.23)\times 10^{-3},\;\;\;
\end{eqnarray}
from which we obtain $|\lambda_{sb}|<0.3\times 10^{-3}$ but the
phase is less constrained again. The corresponding constraint on
Wilson coefficients are
\begin{eqnarray}
 |\Delta C_9| =|C_9-C_9^{SM}|<0.2,\;\;\;
 |\Delta C_{10}| =|C_{10}-C_{10}^{SM}|<2.8.
\end{eqnarray}

Turning to family nonuniveral $Z'$ model in which the coupling
between $Z'$ and a lepton pair is unknown, the two Wilson
coefficients, $C_9$ and $C_{10}$, can be chosen as independent
parameters. Assuming
$\Delta C_9$ and $\Delta C_{10}$ as real, we find
\begin{eqnarray}
 \Delta C_9=0.88\pm0.75,\;\;\; \Delta C_{10}=0.01\pm0.69.
\end{eqnarray}
Removal of the above assumption
leads to
\begin{eqnarray}
 \Delta C_9=-0.81\pm1.22+(3.05\pm0.92)i,\;\;\; \Delta
 C_{10}=1.00\pm1.28+(-3.16\pm0.94)i.
\end{eqnarray}

\begin{figure}\begin{center}
\
\includegraphics[scale=0.8]{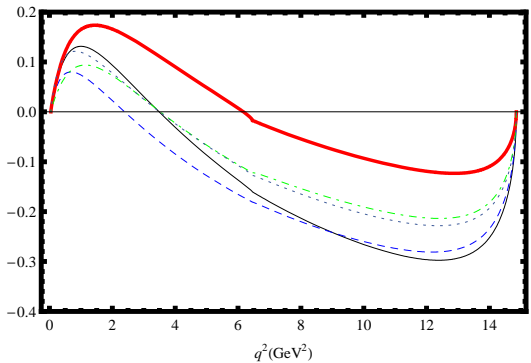}
\caption{Impacts of the NP contributions on normalized
forward-backward asymmetry of $B\to K_2^*l^+l^-$. Black (solid)
line denotes the SM result, while the dashed (blue) and thick (red)
lines correspond to the modification of $C_9$. Dot-dashed
(green) and dotted lines are obtained by modifying $C_{10}$.
}\label{Fig:B-K2-NP-C9-C10}
\end{center}
\end{figure}

 For
illustration, we choose $\Delta C_9=3e^{i\pi/4,i3\pi/4}$ and $\Delta
C_{10}=3e^{i\pi/4,i3\pi/4}$ as the reference points and give the
plots of  FBAs     in Fig.~\ref{Fig:B-K2-NP-C9-C10}. The black
(solid) line denotes the SM result, while the dashed (blue) and
thick (red) lines correspond to the modification of $C_9$. The
dot-dashed (green) and dotted lines are obtained by modifying
$C_{10}$. From the figure for $A_{FB}$, we can see that the
zero-crossing point  $s_0$ can be sizably changed, which can be
tested on the future collider or can be further constrained.

\section{Summary }

Heavy flavor physics has entered a precision era as  large samples of flavor physics data have been
brought to us from $B$ factories and the LHC. As a result, we are able to reach a multitude of
observables from exclusive $b\to s l^+l^-$ processes, which allow to map out the structure of the underlying physics.

In this talk we have concentrated on the $B\to K_1(K^*_0, K_{2}^*, K_3^*, K_4^*) l^+l^-$ in the standard model.
Their branching ratios are predicted to have the order $10^{-6}$ or $10^{-7}$ which are large enough for observation of these processes.  Using the experimental data on $B\to K^*l^+l^-$, we have also presented an update of the constraints on new physics parameters in two specific scenarios and elaborated on the impact on $B\to K_Jl^+l^-$.   We expect more and more results from the LHCb quite soon, which may lead to success of  the justification of new physics degree of freedoms from flavor physics.

 \section*{Acknowledgments}

 We are very grateful to M.Jamil Aslam,
 Run-Hui Li and Yu-Ming Wang for collaboration.
  This work is partly supported by
the National Science Foundation of China under the Grant
No.11075168.


\section*{References}
\begin{thebibliography}{99}
\bibitem{Ali:1999mm}
  A.~Ali, P.~Ball, L.~T.~Handoko and G.~Hiller,
  Phys.\ Rev.\ D {\bf 61} (2000) 074024
  [hep-ph/9910221].

\bibitem{k0}
M.Jamil Aslam, Cai-Dian Lu, Yu-Ming Wang, Phys. Rev. D79 (2009)
074007 e-Print: arXiv:0902.0432 [hep-ph]

\bibitem{k1}
Run-Hui Li, Cai-Dian Lu,
Wei Wang, Phys. Rev. D79 (2009) 094024 e-Print: arXiv:0902.3291
[hep-ph]


\bibitem{k2}
Run-Hui Li, Cai-Dian Lu, Wei Wang,   Phys. Rev. D83 (2011) 034034
e-Print: arXiv:1012.2129 [hep-ph]

\bibitem{kj}
Cai-Dian Lu, Wei Wang,  Phys. Rev. D85 (2012) 034014, e-Print:
arXiv:1111.1513 [hep-ph]


\bibitem{zprime}
Cheng-Wei Chiang, Run-Hui Li, Cai-Dian Lu,  Chin. Phys. C36 (2012)
14-24 e-Print: arXiv:0911.2399 [hep-ph]


\bibitem{Aaij:2011aa}
  RAaij {\it et al.}  [LHCb Collaboration],
  Phys.\ Rev.\ Lett.\  {\bf 108} (2012) 181806
  [arXiv:1112.3515 [hep-ex]].





\bibitem{hep-ph/9812358}
  J.~Charles, A.~Le Yaouanc, L.~Oliver, O.~Pene and J.~C.~Raynal,
  Phys.\ Rev.\ D\ {\bf 60}, 014001  (1999)
  [hep-ph/9812358].




\bibitem{Wirbel:1985ji}
  M.~Wirbel, B.~Stech, M.~Bauer,
  Z.\ Phys.\  {\bf C29}, 637 (1985).


\bibitem{Hatanaka:2009sj}
  H.~Hatanaka and K.~C.~Yang,
  Phys.\ Rev.\  D {\bf 79}, 114008 (2009)
  [arXiv:0903.1917 [hep-ph]];
  Eur.\ Phys.\ J.\  C {\bf 67}, 149 (2010)
  [arXiv:0907.1496 [hep-ph]].



\bibitem{Li:2008tk}
  R.~-H.~Li, C.~-D.~Lu, W.~Wang and X.~-X.~Wang,
  Phys.\ Rev.\ D {\bf 79} (2009) 014013
  [arXiv:0811.2648 [hep-ph]].

\bibitem{Li:2009tx}
  R.~-H.~Li, C.~-D.~Lu and W.~Wang,
  Phys.\ Rev.\ D {\bf 79} (2009) 034014
  [arXiv:0901.0307 [hep-ph]].

\bibitem{Wang:2010ni}
  W.~Wang,
  Phys.\ Rev.\ D {\bf 83} (2011) 014008
  [arXiv:1008.5326 [hep-ph]].







\end{thebibliography}
\end{document}